\documentclass[fleqn,10pt]{wlscirep}
\usepackage{amsmath}

\usepackage{bm}

\title{Tight $N$-observable uncertainty relations and their experimental demonstrations}

\author[1]{Zhi-Xin Chen}
\author[2]{Hui Wang}
\author[2,3]{Jun-Li Li}
\author[1]{Qiu-Cheng Song}
\author[1,3*]{Cong-Feng Qiao}
\affil[1]{School of Physical Sciences, University of Chinese Academy of Sciences, YuQuan Road 19A, Beijing 100049, China}
\affil[2]{College of Materials Science and Opto-Electronic Technology, University of Chinese Academy of Sciences, YuQuan Road 19A, Beijing 100049, China}
\affil[3]{Key Laboratory of Vacuum Physics, University of Chinese Academy of Sciences, YuQuan Road 19A, Beijing 100049, China}
\affil[*]{To whom correspondence should be addressed; E-mail: qiaocf@ucas.ac.cn}

\begin{abstract}
The uncertainty relation, as one of the fundamental principles of quantum physics, captures the incompatibility of noncommuting observables in the preparation of quantum states. In this work, we derive two strong and universal uncertainty relations for $N (N\ge2)$ observables with discrete and bounded spectra, one in multiplicative form and the other in additive form. To verify their validity, for illustration, we implement in the spin-1/2 system an experiment with single-photon measurement. The experimental results exhibit the validity and robustness of these uncertainty relations, and indicate the existence of stringent lower bounds.
\end{abstract}
\begin{document}

\flushbottom
\maketitle
\thispagestyle{empty}

\section*{Introduction}

The original idea of uncertainty relation, one of the most distinct elements of quantum theory, was first introduced by Heisenberg \cite{heis} for the case of position and momentum, and then generalized mathematically by Kennard \cite{Kennard}, Weyl \cite{Weyl}, Robertson \cite{Robertson}, and Schr\"odinger \cite{schrodinger} for two arbitrary observables.
In the literature, the Heisenberg-Robertson uncertainty relation stands as the most representative one:
\begin{eqnarray}\label{Robertson}
(\Delta A)^2(\Delta B)^2\geq
\frac{1}{4} \left|  \langle\psi|[A,B]|\psi\rangle\right|^2 \; , \label{Robertson1}
\end{eqnarray}
where the uncertainty of an observable $A_i$ is characterized by variance $(\Delta A_i)^2 = \langle (A_i - \langle A_i \rangle )^2 \rangle$; $[A,B]=AB-BA$, and the expectation values of operators are defined over a given state $|\psi\rangle$. The uncertainty relation (\ref{Robertson1}) sets an essential limit on the capability of precisely predicting the measurement results of two incompatible observables simultaneously.

The uncertainty relation keeps on being one of the core issues concerned in quantum theory, due to its profound and broad influence in many aspects, e.g. in entanglement detection \cite{Hofmann,OGuhne}, quantum cryptography \cite{Fuchs,Renes}, quantum nonlocality \cite{Oppenheim,ZhihAhnJia}, quantum steering \cite{Schneeloch,YZZhen}, quantum coherence\cite{ZL,Rastegin}, and so on. 
Further research on the uncertainty relation may bring more potentially beneficial applications of quantum physics.
Recently, many theoretical efforts are paid to its improvement and generalization, i.e. to obtain a stronger \cite{mp,sun,song,mbp,xiao,zhang,MCF,LQ} or state-independent \cite{Huang, L1, L2, Branciard, Schwonnek,qian} lower bound, and to deal with more observables \cite {Kechrimparis,dammeier,song,qin,naihuan,long,chenfei,songqc,Dodonov,CFL} or the relativistic system \cite{FZGF}. Several experimental investigations are also performed to check the corresponding relations \cite{xue,dufei,ourexp1}.

Very recently, D. Mondal {\it et al.} proposed two
uncertainty relations for two incompatible observables $A=\sum_{k=1}^{d}a_{k}|a_{k}\rangle\langle a_{k}|$ and $B=\sum_{k=1}^{d}b_{k}|b_{k}\rangle\langle b_{k}|$, which have been decomposed in their eigenbasis, respectively \cite{mbp}:
\begin{eqnarray}\label{mbppro}
(\Delta A)^2(\Delta B)^2\geq
\left(\sum_{k=1}^{d}\tilde{a}_{k}\tilde{b}_{k}\sqrt{F_{\Psi}^{a_{k}}}\sqrt{F_{\Psi}^{b_{k}}}
\right)^2,
\end{eqnarray}
\begin{eqnarray}\label{mbpsum}
(\Delta A)^2+(\Delta B)^2
\geq \frac{1}{2}\sum_{k=1}^{d}\left(\tilde{a}_{k}\sqrt{F_{\Psi}^{a_{k}}}+
\tilde{b}_k\sqrt{F_{\Psi}^{b_{k}}}\right)^2\ .
\end{eqnarray}
Here, $\tilde{a}_{k}={a}_{k}-\langle A \rangle$ and $\tilde{b}_{k}={b}_{k}-\langle B \rangle$; $F_{\Psi}^{x}=|\langle\Psi|x\rangle|^2$ is the fidelity between $|\Psi\rangle$ and $|x\rangle$; $d$ denotes the dimension of the system state;
$\tilde{a}_{k}\sqrt{F_{\Psi}^{a_{k}}}$ and $\tilde{b}_{k}\sqrt{F_{\Psi}^{b_{k}}}$ are arranged such that $\tilde{a}_{k}\sqrt{F_{\Psi}^{a_{k}}}\leq \tilde{a}_{k+1}\sqrt{F_{\Psi}^{a_{k+1}}}$ and $\tilde{b}_{k}\sqrt{F_{\Psi}^{b_{k}}}\leq \tilde{b}_{k+1}\sqrt{F_{\Psi}^{b_{k+1}}}$.
These two uncertainty relations above produce refined lower bounds in terms of the eigenvalues of observables and the transition probabilities between the eigenstates of observables and the state of system.

To capture the incompatibility of noncommuting observables as tight as possible is one of the most important pursuit in physical research of the uncertainty relation.
Moreover, universality is always the core issue concerned in physics, which means it is extremely valuable to derive some uncertainty relations for $N$-observable $(N\ge2)$.
For example, in addition to the pairwise observables, there generally exist the multi-incompatible-observable sets such as tricomponent vectors of angular momentum \cite{dufei} and three Pauli matrices \cite{ourexp1}. Therefore it is necessary and significant to obtain uncertainty relations for $N (N\ge2)$ observables. We notice that (\ref{mbppro}) and (\ref{mbpsum}) can be further optimized and extended nontrivially to cope with any number of observables.

Following we first derive two $N$-observable $(N\ge2)$ uncertainty relations, one in multiplicative form and the other in additive form of variances, and then perform an experiment with single-photon measurement on triple spin operators in the spin-1/2 system to test their validity.

\section*{Results}

\begin{figure} \centering
	\includegraphics[width=0.5\textwidth]{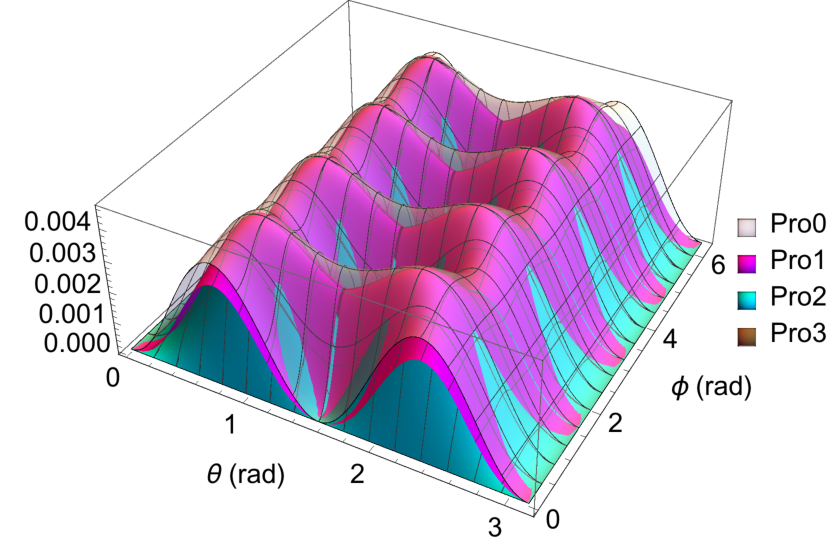}
	\caption{Comparison of the lower bounds in (\ref{pro-hr}), (\ref{pro-fd}), and (\ref{pro-ours}) of the product of variances of three incompatible observables, $S_1$, $S_2$, and $S_3$, the triple spin operators in the spin-1/2 system. The quantum state $|\psi(\theta,\phi)\rangle=\cos\frac{\theta}{2}|+\rangle+e^{i \phi}\sin\frac{\theta}{2}|-\rangle$, with $|+\rangle$ and $|-\rangle$ being the eigenstates of $S_{3}$ corresponding to eigenvalues of $\frac{1}{2}$ and $-\frac{1}{2}$, respectively. The translucent white, magenta, cyan, and brown surfaces, in turn, represent the theoretical values of Pro0, Pro1, Pro2, and Pro3, where Pro0 is the left-hand side (LHS) of relations (\ref{pro-hr}), (\ref{pro-fd}), and (\ref{pro-ours}), and Pro1, Pro2, Pro3 are the RHS of (\ref{pro-ours}), (\ref{pro-fd}), and (\ref{pro-hr}), respectively.
	} \label{profig}
\end{figure}
\vspace{.1cm}

\subsection*{Multiplicative uncertainty relation for $N$ observables}

Given $N$ observables $A_i (i=1,2,...,N)$ in their eigenbasis, i.e., $A_i=\sum_{k=1}^{d} a_{ik}|a_{ik}\rangle \langle a_{ik}|$, with $a_{ik}$ and $|a_{ik}\rangle$, respectively the $k$th eigenvalue and eigenstate of $A_i$, the variances are then $(\Delta A_{i})^2= \sum_{k=1}^{d}\tilde{a}_{ik}^2 \langle |a_{ik} \rangle\langle a_{ik}| \rangle$. Here $\tilde{a}_{ik}={a}_{ik}-\langle A_{i} \rangle$; $\langle |a_{ik}\rangle\langle a_{ik}| \rangle = F_{\psi}^{a_{ik}}$ is the transition probability between the eigenstate of observable and the system state $|\psi \rangle$, or equivalently, the projective probability of $|\psi \rangle$ in the basis $|a_{ik}\rangle$. We set $\vec{u}_i=(u_{i1}, u_{i2}, ...)$=$(|\tilde{a}_{i1}|\sqrt{\langle |a_{i1}\rangle\langle a_{i1}| \rangle},|\tilde{a}_{i2}|\sqrt{\langle |a_{i2}\rangle\langle a_{i2}| \rangle},...)$, which are so arranged such that $u_{ik} \leq u_{i,k+1}$, i.e. $|\tilde{a}_{ik}|\sqrt{\langle |a_{ik}\rangle\langle a_{ik}| \rangle} \leq |\tilde{a}_{i,k+1}|\sqrt{\langle |a_{i,k+1}\rangle\langle a_{i,k+1}| \rangle}$, and hence $ (\Delta A_{i})^2=\sum_{k=1}^{d}u_{ik}^2=\|\vec{u}_i\|^{2} $.
Note that the arranged quantities here are different from those in (\ref{mbppro}) and (\ref{mbpsum}) due to the existence of absolute value. 

By virtue of the Carlson's inequality $\prod_{i=1}^{N}(\sum_{k=1}^{d}u_{ik}^2) \geq [\sum_{k=1}^{d}(\prod_{i=1}^{N}u_{ik}^2)^{1/N}]^N $ \cite{Carlson} which links the arithmetic mean of the geometric means with the geometric mean of the arithmetic means, we obtain the uncertainty relation for $N (N\ge2)$ observables in product form:
\begin{eqnarray}\label{ur-pro}
\prod_{i=1}^{N}\left(\Delta A_i \right)^2 \geq \left[\sum_{k=1}^{d}\left(\prod_{i=1}^{N}|\tilde{a}_{ik}|^2 \langle |a_{ik}\rangle\langle a_{ik}| \rangle \right)^\frac{1}{N}\right]^N\ .
\end{eqnarray}
Here, the lower bound is formulated in terms of the eigenvalues of observables and the transition probabilities between the eigenstates and the state of system. 
It is optimization-free, i.e., independent of any optimization like finding some orthogonal states to maximize the lower bound.
The uncertainty relation (\ref{ur-pro}) is universal for any number of observables while many uncertainty relations proposed before are only available for $N=2$ or $N\ge3$. It is strong as well, e.g., the lower bound of (\ref{ur-pro}) is tighter than (\ref{mbppro}) when $N=2$, since
\begin{eqnarray}\label{prove}
\left(\sum_{k=1}^{d} v_{1k} \cdot v_{2k} \right)^2 \le \left(\sum_{k=1}^{d} |v_{1k}| \cdot |v_{2k}| \right)^2\nonumber \le \left(\sum_{k=1}^{d} |v_{1k}|_\uparrow \cdot |v_{2k}|_\uparrow \right)^2\ .
\end{eqnarray}
Here, $v_{ik}
=\tilde{a}_{ik}\sqrt{\langle |a_{ik}\rangle\langle a_{ik}|\rangle}$,
and $ |v_{ik}|_\uparrow$ is the increasing sequence of $|v_{ik}| $.
Evidently, $(\sum_{k=1}^{d} v_{1k} \cdot v_{2k} )^2$ is the right-hand side (RHS) of (\ref{mbppro}), while $(\sum_{k=1}^{d} |v_{1k}|_\uparrow \cdot |v_{2k}|_\uparrow )^2$ is the RHS of (\ref{ur-pro}) when $N=2$. Hence, the new bound of (\ref{ur-pro}) is stronger than (\ref{mbppro}).

\begin{figure} \centering
	\includegraphics[width=0.5\textwidth]{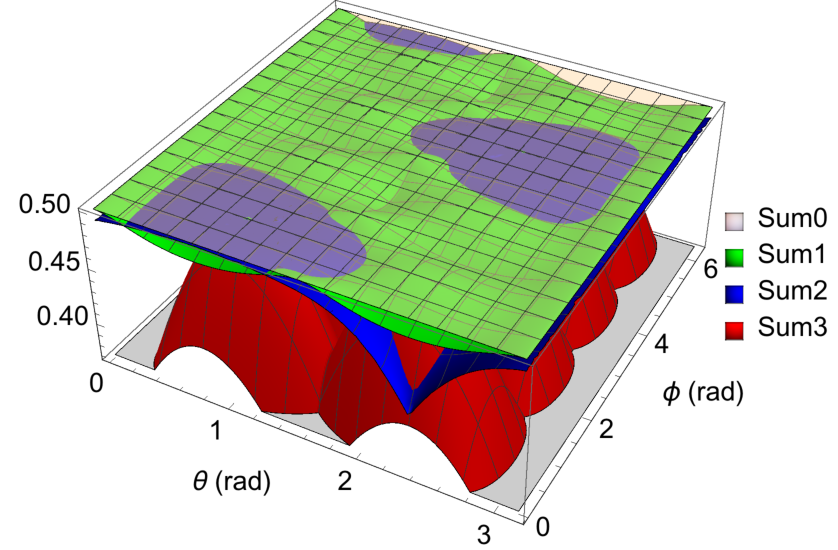}
	\caption{Comparison of the lower bounds in (\ref{sum-ours}), (\ref{sum-song}), and (\ref{sum-fd}) of the sum of variances of three incompatible observables, $S_1$, $S_2$, and $S_3$, the triple spin operators in the spin-1/2 system. The quantum state $|\psi(\theta,\phi)\rangle=\cos\frac{\theta}{2}|+\rangle+e^{i \phi} \sin\frac{\theta}{2} |-\rangle$, with $|+\rangle$ and $|-\rangle$ being the eigenstates of $S_{3}$ corresponding to eigenvalues of $\frac{1}{2}$ and $-\frac{1}{2}$, respectively. The translucent white, green, blue, and red surfaces, in turn, represent the theoretical values of Sum0, Sum1, Sum2, and Sum3, where Sum0 is the LHS of relations (\ref{sum-ours}), (\ref{sum-song}), and (\ref{sum-fd}), and Sum1, Sum2, Sum3 are the RHS of (\ref{sum-ours}), (\ref{sum-song}), and (\ref{sum-fd}), respectively.} \label{sumfig}
\end{figure}
\vspace{.1cm}

\begin{figure} \centering
	\includegraphics[width=0.9\textwidth]{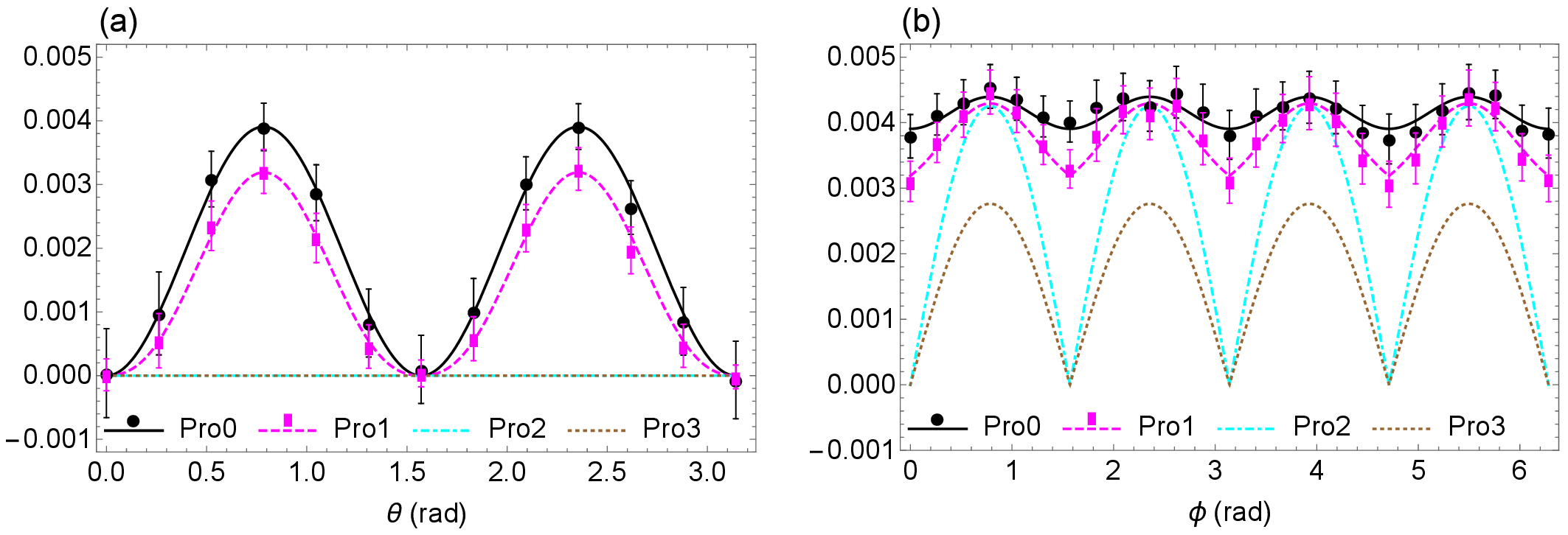}
	\caption{Experimental results of multiplicative uncertainty relations. (a) Experimental results for $ S_i (i=1,2,3)$, the triple components of the angular momentum in the spin-1/2 system with the states $|\psi(\theta,0)\rangle = \cos(\theta/2) |+\rangle + \sin(\theta/2)|-\rangle$.
	(b) Experimental results for $ S_i (i=1,2,3)$, with the quantum states  $|\psi(\pi/4,\phi)\rangle = \sqrt{2+\sqrt{2}}/{2} |+\rangle + e^{i\phi} \sqrt{2-\sqrt{2}}/{2}|-\rangle$. In both (a) and (b), the solid black, dashed magenta, dot-dashed cyan, and dotted brown curves, in turn, represent the theoretical values of Pro0, Pro1, Pro2, and Pro3, where Pro0 is the LHS of relations (\ref{pro-hr}), (\ref{pro-fd}), and (\ref{pro-ours}), and Pro1, Pro2, Pro3, in turn, are the RHS of (\ref{pro-ours}), (\ref{pro-fd}), and (\ref{pro-hr}). The black circles and magenta rectangles represent the experimental values of Pro0 and Pro1, respectively. Error bars represent $\pm1$ standard deviation.
	} \label{proexp}
\end{figure}

In the literature, there exist uncertainty relations for three incompatible observables \cite{dufei}. For comparison, specifically in the spin-1/2 system $(\hbar=1)$ with three angular momentum operators $S_1$, $S_2$, and $S_3$, they write
\begin{equation}\label{pro-hr}
\prod_{i=1}^{3}(\Delta S_i)^2  \ge \frac{1}{8}
\left| {\langle {S_1}\rangle }  {\langle {S_2}\rangle }  {\langle {S_3}\rangle } \right|\ ,
\end{equation}
\begin{equation}\label{pro-fd}
\prod_{i=1}^{3}(\Delta S_i)^2  \ge \frac{1}{3\sqrt{3}}
\left| {\langle {S_1}\rangle \langle {S_2}\rangle \langle {S_3}\rangle } \right|\ .
\end{equation}
Here, the operators satisfy the commutation relations $[S_1,S_2]=i S_3$, $[S_2,S_3] = i S_1$, and $[S_3,S_1]=i S_2$.
In the representation of $S_1$, $S_2$, and $S_3$, the uncertainty relation (\ref{ur-pro}) turns to
\begin{flalign}\label{pro-ours}
\prod_{i=1}^{3}(\Delta S_i)^2
\geq \left[\sum_{k=1}^{2}\left(\prod_{i=1}^{3}\tilde{s}_{ik}^2
\langle |s_{ik}\rangle\langle s_{ik}| \rangle \right)^\frac{1}{3}\right]^3 =\left[\sum_{k=1}^{2}\left( \frac{1}{8}\prod_{i=1}^{3} \left(1+(-1)^k | \langle S_i \rangle | \right)^2 \left(1-(-1)^k | \langle S_i \rangle | \right) \right)^{\frac{1}{3}}\right]^3
\end{flalign}
with $S_i=\sum_{k=1}^{d}s_{ik}|s_{ik}\rangle\langle s_{ik}|$ and $\tilde{s}_{ik} = {s}_{ik}-\langle S_{i} \rangle$.
Note, the uncertainty relations (\ref{pro-hr}) and (\ref{pro-fd}) suffer the triviality problem, i.e., they will become to nought when any of the expectation values is zero as shown in Fig. \ref{proexp}(a), while (\ref{pro-ours}) has no such problem.
Evidently, the uncertainty relation (\ref{pro-fd}) is tighter than (\ref{pro-hr}) which is a straightforward extension of (\ref{Robertson1}).

As shown in Fig. \ref{profig}, the uncertainty relation (\ref{pro-ours}) is superior in lower bound to (\ref{pro-hr}) and (\ref{pro-fd}) in most of the kinematic region, especially as shown in the case of triple incompatible observables $ S_i (i=1,2,3)$ and when the quantum state of a spin-1/2 system is parameterized by $\theta$ and $\phi$ as
\begin{align}\label{state}
|\psi(\theta,\phi)\rangle=\cos\frac{\theta}{2}|+\rangle+e^{i \phi}\sin\frac{\theta}{2}|-\rangle\ .
\end{align}
Here, $|+\rangle$ and $|-\rangle$ are the eigenstates of $S_{3}$ corresponding to eigenvalues of $\frac{1}{2}$ and $-\frac{1}{2}$, respectively; $\theta\in[0,\pi]$ and $\phi\in[0,2\pi]$. Note that $|\psi(\theta,\phi)\rangle$ represents any pure state of spin-1/2 quanta on the surface of the Bloch sphere.

\begin{figure}\centering
	\includegraphics[width=0.5\textwidth]{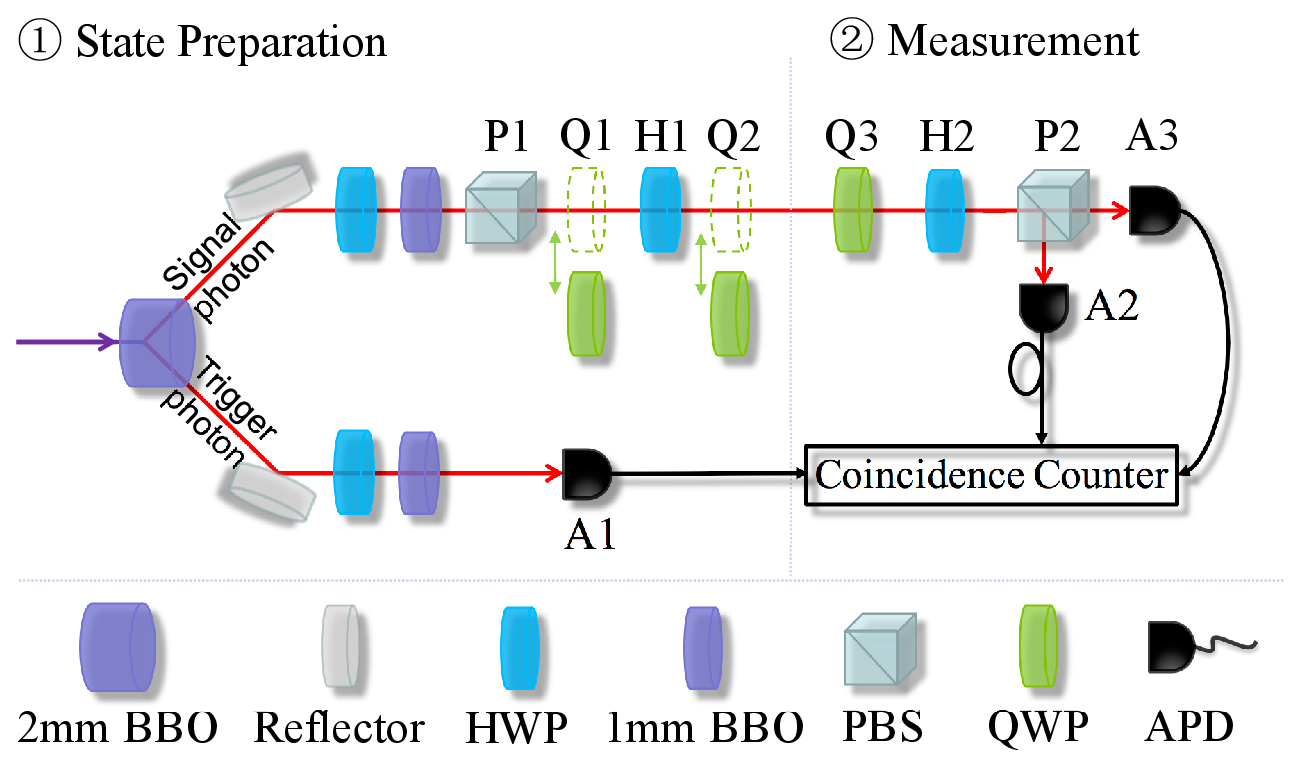}
	\caption{Experimental setup. In the stage of state preparation,
		the photon pair is generated due to the type-II spontaneous parametric down-conversion (SPDC) in the 2mm BBO.
		The trigger photon in the lower path goes to the single-photon avalanche photodiode (APD, A1), and the signal photon in the upper path is heralded and then initialized to the state $|+\rangle$ by the polarizing beam splitter (PBS, P1). The quantum states $ |\psi(\theta,\phi)\rangle $ are prepared by the combined operations of a quarter-wave plate (QWP, Q1), a half-wave plate (HWP, H1) and Q2. In the stage of measurement, Q3, H2 and P2 are used jointly to realize the projective measurements in different bases, or equivalently, the measurements of expectation of $S_i (i=1,2,3)$. A1, A2 and A3 are all connecting to the coincidence counter.
	}\label{f1}
\end{figure}

\subsection*{Additive uncertainty relation for $N$ observables}

It is well-known that the multiplicative uncertainty relation may become trivial when the state of the system happens to be the eigenstate of one of the observables. To avoid this kind of triviality, constructing the uncertainty relation in the form of summation is definitely necessary and valuable. 

For $N\ge2$, one has a universal inequality
\begin{equation}\label{mp}
\sum_{i=1}^{N}\|\vec{u}_{i}\|^{2}\geq \frac{1}{2N-2} \sum_{1\leq i<j\leq N} \|\vec{u}_{i}+\vec{u}_{j}\|^{2}.
\end{equation}
While for $N\ge3$, there is a stringent inequality \cite{chenfei} for multiple vectors
\begin{equation}\label{cf}
\sum_{i=1}^{N}\|\vec{u}_{i}\|^{2}\geq\frac{1}{N-2}\left[\sum_{1\leq i<j\leq N} \|\vec{u}_{i}+\vec{u}_{j}\|^{2} -\frac{1}{(N-1)^{2}}\left(\sum_{1\leq i<j\leq N}\|\vec{u}_{i}+\vec{u}_{j}\|\right)^{2}\right] .
\end{equation}
It has been proven that the RHS of (\ref{cf}) is tighter than the RHS of (\ref{mp}) when $N\ge3$. However, (\ref{cf}) is not available for $N=2$ since the denominator $N-2$.

Combining the universality of (\ref{mp}) for $N=2$ and the stringency of (\ref{cf}) for $N\ge3$, we derive the following strong inequality for any number of vectors,  
\begin{equation}\label{sumine}
\begin{split}
\sum_{i=1}^{N}\|\vec{u}_{i}\|^{2}
\geq\frac{1}{2^{\bm{H}(2-N)}N-2}\left[\sum_{1\leq i<j\leq N} \|\vec{u}_{i}+\vec{u}_{j}\|^{2} +\frac{\bm{H}(2-N)-1}{(N-1)^{2}}\left(\sum_{1\leq i<j\leq N}\|\vec{u}_{i}+\vec{u}_{j}\|\right)^{2}\right] 
\end{split}
\end{equation}
where $\bm{H}(x)$ is the unit step function whose value is zero for $x<0$ and one for $x \ge 0$.
Employing this mathematical inequality with the same definition for $N$ observables $A_i$ in above, we obtain the additive uncertainty relation for $N$ observables
\begin{equation}\label{ur-sum}
\begin{split}
\sum_{i=1}^{N}(\Delta A_{i})^{2}\geq \frac{1}{2^{\bm{H}(2-N)}N-2}\left[\sum_{1\leq i<j\leq N}\Lambda_{ij}^{2} + \frac{\bm{H}(2-N)-1}{(N-1)^{2}}\left(\sum_{1\leq i<j\leq N}\Lambda_{ij}\right)^{2} \right]\ ,
\end{split}
\end{equation}
where
\begin{eqnarray}\label{}
\Lambda_{ij}^{2}&=& \sum_{k=1}^{d}\left(|\tilde{a}_{ik}|\sqrt{\langle |a_{ik}\rangle \langle a_{ik}| \rangle} + |\tilde{a}_{jk}| \sqrt{\langle |a_{jk} \rangle\langle a_{jk}| \rangle}\right)^2 . \notag
\end{eqnarray}

While most uncertainty relations proposed before are only available for pairwise or multiple observables, the uncertainty relation (\ref{ur-sum}) is universal for any number of observables. It is optimization-free and stringent as well. Since
\begin{eqnarray}\label{sum-prove}
\Lambda_{ij}^{2}
=\sum_{k=1}^{d}\left( |v_{ik}|_\uparrow + |v_{jk}|_\uparrow \right)^2
\ge \sum_{k=1}^{d}\left( v_{ik} + v_{jk} \right)^2\ ,
\end{eqnarray}
we can prove that when $N=2$, the RHS of (\ref{ur-sum}), $\frac{1}{2}\Lambda_{12}^{2}$, is tighter than the RHS of (\ref{mbpsum}), $\frac{1}{2} \sum_{k=1}^{d}\left( v_{1k} + v_{2k} \right)^2$.
For $N\ge 3$, it is obvious that (\ref{ur-sum}) is stronger than the simple generalization of (\ref{mbpsum}) because the RHS of (\ref{cf}) is tighter than the RHS of (\ref{mp}). 
The lower bound of (\ref{ur-sum}) is formulated by the eigenvalues of observables and the transition probabilities between the eigenstates and the system state.
It is different from the existing muti-observable uncertainty relations constructed in terms of variances or standard deviations, like \cite{songqc}
\begin{align}\label{song}
\sum_{i=1}^{N}(\Delta A_{i})^{2}
\geq\frac{1}{N}\left[\Delta\left(\sum_{i=1}^{N}A_{i}\right)\right]^{2} +\frac{2}{N^2(N-1)}\left[\sum_{1\leq i<j\leq N}\Delta(A_{i}-A_{j})\right]^{2}\ 
\end{align}
which is more stringent in qubit systems than the relation obtained directly from (\ref{cf}) and constructed in terms of variances.

In comparison with other summed forms of uncertainty relation, we take $N=3$ and $ S_i (i=1,2,3)$, the operators of angular momentum in the spin-1/2 system as an example. In this case, the uncertainty relation (\ref{ur-sum}) now turns to
\begin{eqnarray}\label{sum-ours}
\sum_{i=1}^{3}(\Delta S_i)^2 \geq
\sum_{1\leq i<j\leq 3}\Omega_{ij}^{2}-\frac{1}{4}\left(\sum_{1\leq i<j\leq 3}\Omega_{ij}\right)^{2}
\end{eqnarray}
with
\begin{eqnarray}
\Omega_{ij}^{2}= \sum_{k=1}^{2} \left(|\tilde{s}_{ik}|\sqrt{\langle |s_{ik}\rangle\langle s_{ik}| \rangle}+|\tilde{s}_{jk}|\sqrt{\langle |s_{jk}\rangle\langle s_{jk}| \rangle}\right)^2\ .\notag
\end{eqnarray}
And the uncertainty relation (\ref{song}) simplifies to
\begin{align}\label{sum-song}
\sum_{i=1}^{3}(\Delta S_{i})^{2}
\geq\frac{1}{3}\left[\Delta\left(\sum_{i=1}^{3}S_{i}\right)\right]^{2}+\frac{1}{9}\left[\sum_{1\leq i<j\leq 3}\Delta(S_{i}-S_{j})\right]^{2}.
\end{align}

In the literature, there exists another kind of uncertainty relation in summed form for $ S_i (i=1,2,3)$ \cite{dufei}, a special case of (26) in the reference \cite{song}, which has been testified in an experiment with a negatively charged nitrogen-vacancy centre in diamond, i.e.,
\begin{equation}\label{sum-fd}
\begin{aligned}
\sum_{i=1}^{3}(\Delta S_i)^2 \ge \frac{1}{\sqrt{3}}
\left( {\left| {\langle {S_1}\rangle } \right| + \left| {\langle {S_2}\rangle } \right| + \left| {\langle {S_3}\rangle } \right|} \right)\ .
\end{aligned}
\end{equation}

As illustrated in Fig. \ref{sumfig}, the lower bound of the new additive uncertainty relation (\ref{sum-ours}) is more stringent than those given by (\ref{sum-song}) and (\ref{sum-fd}) in most cases, with the quantum state of the system taken to be the typical one of (\ref{state}).

\section*{Experimental demonstrations}

\begin{figure} \centering
	\includegraphics[width=0.9\textwidth]{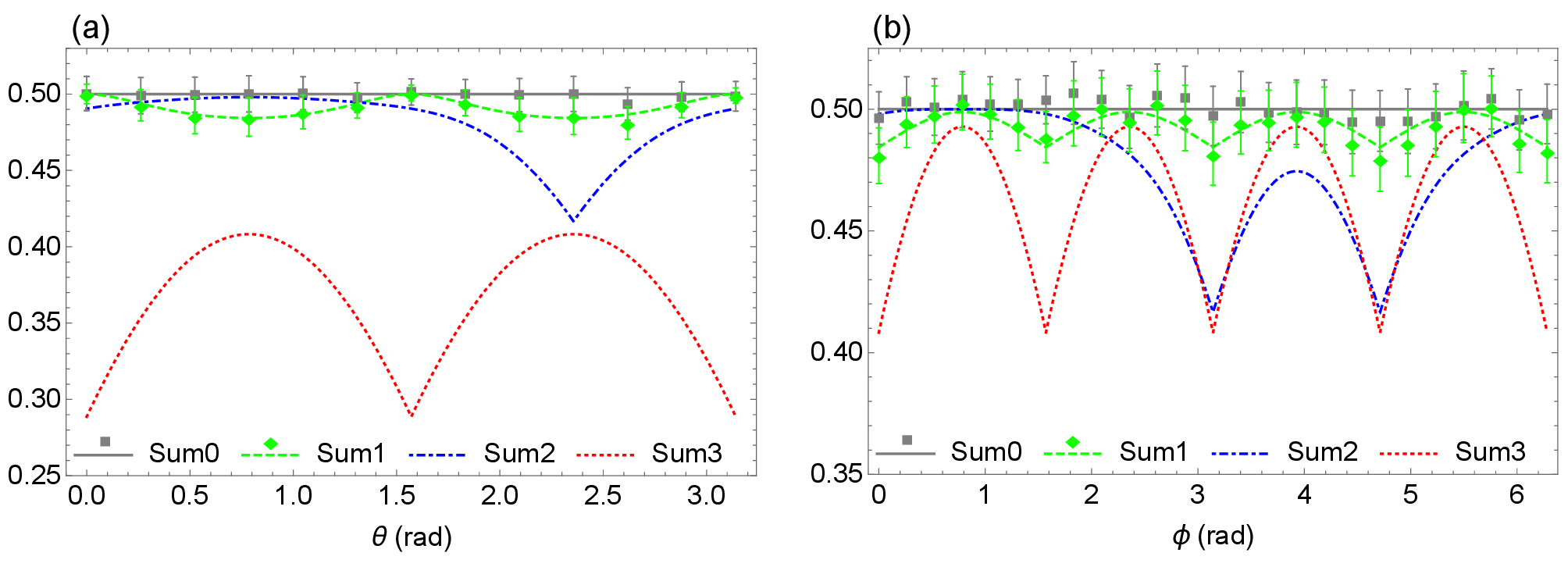}
	\caption{Experimental results of additive uncertainty relations. (a) Experimental results for $ S_i (i=1,2,3)$, the triple components of the angular momentum in the spin-1/2 system with the states $|\psi(\theta,0)\rangle = \cos(\theta/2) |+\rangle + \sin(\theta/2)|-\rangle$. (b) Experimental results for $ S_i (i=1,2,3)$, with the quantum states $|\psi(\pi/4,\phi)\rangle = \sqrt{2+\sqrt{2}}/{2} |+\rangle + e^{i\phi} \sqrt{2-\sqrt{2}}/{2}|-\rangle$. In both (a) and (b), the solid gray, dashed green, dot-dashed blue, and dotted red lines, in turn, represent the theoretical values of Sum0, Sum1, Sum2, and Sum3, where Sum0 is the LHS of relations (\ref{sum-ours}), (\ref{sum-song}), and (\ref{sum-fd}), and Sum1, Sum2, Sum3 are in turn the RHS of (\ref{sum-ours}), (\ref{sum-song}), and (\ref{sum-fd}). The gray squares and green diamonds represent the experimental values of Sum0 and Sum1, respectively.
	}\label{sumexp}
\end{figure}

To testify the uncertainty relations (\ref{ur-pro}) and (\ref{ur-sum}), more specifically (\ref{pro-ours}) and (\ref{sum-ours}), we implement the experiment with single-photon measurement which is convenient and reliable.
As shown in Fig. \ref{f1}, two main stages are undergone in the experimental setup, i.e., the state preparation and projective measurement of the quantum system. Here the spin-1/2 system, the qubit $ |\psi(\theta,\phi)\rangle $, is constructed by the polarized states of a single photon extracting from a pair of photons by triggering one of them. The horizontal polarization $ |H\rangle $ and vertical polarization $ | V \rangle $ of the photon are represented by  $|+\rangle$ and $|-\rangle$, respectively.

In the stage of state preparation, we use a continuous wave diode laser
with wavelength 405 nm to pump the 2-mm-thick nonlinear barium borate (BBO) crystal. Due to the effect of type-\uppercase\expandafter{\romannumeral2} spontaneous parametric down-conversion (SPDC), the photon pair at a wavelength of 810 nm is produced in the 2-mm BBO. Using one half-wave plate (HWP) and one 1-mm-thick BBO in each path to compensate the birefringent walk-off effect in the main BBO(2mm). After the detection of a trigger photon
by the first single-photon avalanche photodiode (APD, A1) in Fig. \ref{f1}, the signal photon
is heralded and initialized to the state $|+\rangle$ by the first polarizing beam splitter (PBS, P1). Then we use the combined operation of a quarter-wave plate (QWP, Q1), H1 and Q2 to generate the interested quantum state $ |\psi(\theta,\phi)\rangle $. Note that Q1 and Q2 are removed during the preparation of $ |\psi(\theta,0)\rangle $, and inserted while preparing the state $ |\psi(\pi/4,\phi)\rangle $.
In practice, we prepare two series of quantum states, i.e., $ |\psi(\theta,0)\rangle , \theta=n \pi/12$ $(n =0, 1, . . ., 12) $, and $ |\psi(\pi/4,\phi)\rangle , \phi=n\pi/12$  $(n =0, 1, . . . ,24)$.

In the stage of measurement, we use Q3, H2 and P2 jointly to realize the projective measurement on different bases, i.e. $\langle |s_{ik}\rangle\langle s_{ik}| \rangle = F_{\psi}^{s_{ik}}$ , or equivalently the measurement of expectation values of $S_1$, $S_2$ and $S_3$ . Finally, the coincidence counter, connected to three APDs, A1, A2, and A3, outputs the coincidence measurement on the trigger-signal photon pair. In the end, the coincidence counting rate counts about 2800 $s^{-1}$.

Figure. \ref{proexp} illustrates the experimental measurement of multiplicative uncertainty relation (\ref{ur-pro}) or specifically (\ref{pro-ours}), in contrast with the other two lower bounds of (\ref{pro-hr}) and (\ref{pro-fd}). The states $|\psi(\theta,0)\rangle$ and $|\psi(\frac{\pi}{4}, \phi)\rangle$ are employed for Figs. \ref{proexp}(a) and \ref{proexp}(b), respectively.
Note that the lower bounds of (\ref{pro-hr}) and (\ref{pro-fd}) become trivial due to $\langle S_2 \rangle =0 $ in Fig. \ref{proexp}(a), while the new lower bound (\ref{pro-ours}) still imposes a strong restriction on the product of variances. The results, which fit the theoretical predictions well, show that for the product $\prod_{i=1}^{3}(\Delta S_i)^2 $, the lower bound in (\ref{pro-ours}) is more stringent than those of (\ref{pro-hr}) and (\ref{pro-fd}).

In Fig. \ref{sumexp}, we testify the additive uncertainty relation (\ref{ur-sum}), and show the relative stringency of different lower bounds of (\ref{sum-ours}),  (\ref{sum-song}), and (\ref{sum-fd}) using the states $|\psi(\theta,0)\rangle$ and $|\psi(\frac{\pi}{4}, \phi)\rangle$ for Figs. \ref{sumexp}(a) and \ref{sumexp}(b), respectively. The experimental results, fitting the predictions well, indicate that for sum $\sum_{i=1}^{3}(\Delta S_i)^2 $ the lower bound of (\ref{sum-ours}) is more stringent than those of (\ref{sum-song}) and (\ref{sum-fd}).

In the above figures, the experimental error, standing for the $\pm1$ standard deviation, mainly comes from the fluctuation of photon counting due to the instability of laser power and the probabilistic SPDC, and the imperfection of experiment devices, such as wave plates and APDs.

\section*{Conclusion}

To conclude, we derive two tight and universal uncertainty relations for $N (N\ge2)$ observables, one in multiplicative form and the other in additive form of variances.
The measure taken in deriving the new inequalities brings new insight to the study of the relationship between uncertainty and non-orthogonality for $N$ observables.
In comparison with other uncertainty relations, it is found that the results given in this work are generally more stringent in lower bound than others. We also implement a practical experiment with single-photon measurement to testify the theoretical predictions, especially for
spin-1/2, and find that the new uncertainty relations are valid and superior. 
Notice that the tighter uncertainty relations with experimental testification are important not only for a better understanding of the foundation of quantum theory, but also for the
quantum information applications like
the enhancement of precise quantum measurement.

\section*{Acknowledgements}
This work was supported, in part, by the Ministry of Science and Technology of the People's Republic of China (2015CB856703); by the Strategic Priority Research Program of the Chinese Academy of Sciences, Grant No. XDB23030100; and by the National Natural Science Foundation of China (NSFC) under the Grants No. 11375200 and No. 11635009.

\section*{Contributions}
Z.-X. Chen and C.-F. Qiao proposed the project and established the main results. 
Z.-X. Chen, H. Wang and Q.-C. Song performed the calculations and plotted the figures.
Z.-X. Chen, J.-L. Li, and C.-F. Qiao designed the experiment.
Z.-X. Chen, H. Wang, and C.-F. Qiao performed the experiment.
Q.-C. Song and J.-L. Li joined discussions and provided suggestions. 
All authors analysed the results and wrote the manuscript.

\section*{Competing Interests}
The authors declare no competing interests.

\end{document}